\documentclass{jpconf}
\usepackage{epsfig}
\usepackage{amsmath}
\usepackage{mathrsfs}
\usepackage{subfigure}
\usepackage[square,sort&compress]{natbib}
\usepackage{rotating}

\newcommand{\be}{\begin{equation}}
\newcommand{\ee}{\end{equation}}

\def\ltsima{$\; \buildrel < \over \sim \;$}
\def\simlt{\lower.5ex\hbox{\ltsima}}
\def\gtsima{$\; \buildrel > \over \sim \;$}
\def\simgt{\lower.5ex\hbox{\gtsima}}
\newcommand\sgra{Sgr~A$^*$}

\def\del#1{{}}

\def\msun{{\,{\rm M}_\odot}}

\begin{document}

\title{Simulations of the formation of a gaseous disc and young stars near Sgr A*
via cloud--cloud collisions}

\author[Sergei Nayakshin and Alexander Hobbs]{Sergei Nayakshin and Alexander Hobbs}

\address{Dept. of Physics \& Astronomy, University of Leicester, Leicester, LE1 7RH, UK}

\ead{sergei.nayakshin@astro.le.ac.uk}

\begin{abstract}
Young massive stars in the central parsec of our Galaxy are best explained by
star formation within at least one, and possibly two, massive self-gravitating
gaseous discs. With help of numerical simulations, we here consider whether
the observed population of young stars could have originated from a large
angle collision of two massive gaseous clouds at $R \simeq 1$ pc from \sgra.
In all the simulations performed, the post-collision gas flow forms an inner
nearly circular gaseous disc and one or two eccentric outer filaments,
consistent with the observations. Furthermore, the radial stellar mass
distribution is always very steep, $\Sigma_* \propto R^{-2}$, again consistent
with the data. The 3D velocity structure of the stellar distribution is however
sensitive to initial conditions (e.g., the impact parameter of the clouds) and
gas cooling details. In all the cases the amount of gas accreted by our inner
boundary condition is large, enough to allow \sgra to radiate near its
Eddington limit during $\sim 10^5$ years. This suggests that a refined model
would have physically larger clouds (or a cloud and a disc such as the CND)
colliding at a distance of a few pc rather than 1 pc as in our simulations.
\end{abstract}

\section{Introduction}

The central parsec of our Galaxy is host to young, massive ``He-I'' stars that
can be classified into two stellar discs that orbit the central
supermassive black hole, \sgra, outside the inner arcsecond. Most of the stars are located in a
well-defined thin stellar disc that rotates clockwise as seen on the sky
\citep{Levin03,GenzelEtal03,PaumardEtal06}. The rest of the stars can be
classified as a second more diffuse disc that is more eccentric and rotating counter clockwise
\citep{GenzelEtal03,PaumardEtal06} although the statistical significance of
this feature is disputed by \cite{LuEtal06}. The stellar populations are
relatively co-eval, formed approximately $6 \pm 1$ million years ago \citep{PaumardEtal06}. In terms of
formation scenarios, the ``in-situ'' model is currently
the one best favoured by the Galactic Centre (GC) community; in this picture,
stars form inside a self-gravitating gaseous disc
\citep[e.g.,][]{Levin03,NC05,PaumardEtal06}. Theorists had
expected gaseous massive discs around supermassive black
holes to form stars or planets
\citep[e.g.,][]{Paczynski78,Goodman03}
long before the properties of the young stars in the GC became
known. Recently, \cite{NayakshinEtal07} have numerically simulated the
fragmentation process of a geometrically thin gaseous disc of mass $\simgt
10^4$ Solar mass for conditions appropriate for our GC (albeit with a rather
simple cooling prescription) and found a top-heavy mass function for the stars
formed there. \cite{AlexanderEtAl08} extended these numerical studies to
the fragmentation of eccentric accretion disks.

However, the in-situ model has not so far addressed in detail the origin of
the gaseous discs themselves. Similar discs are believed to exist in AGN and
quasars, and indeed are invoked as a means of feeding a supermassive black
hole's immense appetite for gaseous fuel \citep[e.g.][]{Frank02}. However,
\sgra is currently not a member of the AGN club, as its bolometric luminosity is around
$\sim 10^{-9}$ of its Eddington limit \citep[e.g.,][]{Baganoff03a}, implying
a quiescent mode of accretion. 

The viscous timescale of a thin, marginally self-gravitating disc around
\sgra, given a disc mass of $\approx 10^4 \msun$ \citep[reasons for this
choice of mass see in, e.g.,][]{NayakshinEtal06} is very long compared with the age of the stellar systems
(approx. $10^9$ yrs compared to $10^6$ yrs). In fact, the gaseous discs
would have to evolve even faster as the two stellar systems are largely co-eval. The requisite gaseous discs
could therefore not have been assembled by viscous transport of angular momentum.

There is also no strong evidence for other similar star formation events in
the region within the last $\sim 10^8$ years. Therefore, the one-off star
formation event appears to be best explained by a one-off deposition of gas
within the central parsec. There are several ways in which this could have
happened, e.g. a Giant Molecular Cloud (GMC) with a sub-parsec impact
parameter (in relation to \sgra) could have self-collided and become partially
bound to the central parsec \citep[e.g.,][]{NC05}. Alternatively, a GMC could
have struck the Circumnuclear Disc (CND) located a few pc away from \sgra, and
then created gas streams that settled into the central parsec.

In this article we explore such a one-off collision event in a very simple
setup. We allow two massive, uniform and spherical clouds on significantly
different orbits to collide with each other one parsec away from \sgra, and
follow the gas dynamics in some detail. We find that the collision forms
streams of gas with varying angular momentum, both in magnitude and
direction. Parts of these streams collide and coalesce to form a disc in the
inner region. As the gas cools, it becomes self-gravitating, and stars are
born. The resulting distribution of stellar orbits is compared
with the observational data throughout the article, and particularly in \S
\ref{sec:discussion}. We find that our model is reasonably successful in explaining the population of young He-I
stars in the Galactic Centre.

\section{Numerical methods}\label{sec:numerics}

The numerical approach and the code used in this article is the same as in
\cite{NayakshinEtal07} with only slight modifications. Namely, we employ
GADGET-2, a smoothed particle hydrodynamic (SPH)/N-body code. Gas cools
according to $du/dt = - u/t_{\rm cool}(R)$, where cooling time depends on
radius as
\begin{equation}
t_{\rm cool}(R) = \beta  t_{dyn}(R)
\label{beta}
\end{equation}
\noindent where $t_{\rm dyn} = 1/\Omega$ and $\Omega = (GM_{\rm
bh}/R^{3})^{1/2}$, the Keplerian angular velocity, where $\beta$ is a
dimensionless number.  Gas moves in the gravitational potential of \sgra,
modelled as a motionless point mass with $M_{\rm bh} = 3.5 \times 10^{6}
\msun$ at the origin of the coordinate system, and a much older relaxed
isotropic stellar cusp potential given by the mass profile derived by
\cite{GenzelEtal03}.

To model star formation, we introduce sink particles when gas density exceeds
a ``critical density'' \citep[see][]{NayakshinEtal07}. Accretion of gas onto stars is calculated using the Bondi-Hoyle rate, which is
capped at Eddington.  Our units are $M_{u} = 3.5 \times 10^{6}
\msun$, the mass of \sgra, $R_{u} = 1.2 \times 10^{17}$ cm $\approx 0.04$ pc,
equal to 1'' when viewed from the $\approx$ 8 kpc distance to the GC, and
$t_{u} = 1/\Omega(R_{u})$, the dynamical time evaluated at $R_{u}$,
approximately 60 years. We use $R$ to signify distance in physical units and
the dimensionless $r = R/R_{u}$ throughout the article inter-exchangeably.

\section{Initial conditions}\label{sec:ic}

Our initial conditions all comprise a collision between two gas clouds at the edge of the inner parsec of
the GC. The specific parameters for each run, labelled S1 to S6, can be found
in Table 1. Each cloud is spherically symmetric and of
uniform density, containing less than 1\% of the mass of \sgra. The clouds are
mainly composed of molecular hydrogen, using $\mu = 2.46$. The initial temperature of the clouds is arbitrarily set to
20K. A more realistic model could have included a turbulent velocity and
density field, but for practical purposes we limit ourselves to a smaller set of input parameters for this first study.

We define the primary to be the larger cloud, with a radius of $R_1 = 0.2$ pc
and $M_{1} = 3.4 \times 10^{4} \msun$, whilst the secondary has a radius of
$R_2 = 0.172$ pc with $M_{2} = 2.6 \times 10^{4} \msun$.  The initial
positions of the cloud centres are the same for all the simulations, and are
$r_{1} = (25,0,0)$ and $r_{2} = (22,6,7)$ in dimensionless units.  The
initial density of the clouds is slightly above the tidal shear density at
their initial locations.

In all the tests the absolute velocity of cloud 1 is of the order of the
Keplerian velocity at that radius (and would be exactly circular in the
absence of the cusp, $M_{\rm cusp} = 0$). Orbits in a cusped potential however
are not Keplerian. The
initial specific energy and angular momentum $l$ of a particle can be
used to find the pericenter, $r_{\rm pe}$,  and the apocenter, $r_{\rm ap}$,
of the orbit. The orbit's eccentricity $e$ is then defined via
\begin{equation}
\frac{1+e}{1-e} = \frac{r_{ap}}{r_{pe}}\;.
\label{ecceq}
\end{equation}
For any hyperbolic orbits, the $e$ is capped at 1. The orbit of
cloud 1 is slightly eccentric, with pericentre and apocentre of 25 and 31.5
respectively, and with $e = 0.12$. The initial orbit trajectory of the second cloud is
varied between the tests to cover a small range of possibilities. The
parameters for this trajectory in terms of pericentre, apocentre and
eccentricity are given in Table \ref{table1}. The collision itself is highly
supersonic.

\begin{table}
\caption{Initial conditions of our simulations. The
meaning of the symbols in the Table are: $\beta$ is the cooling parameter,
$\bf{v_1}$, $\bf{v_2}$ are the initial velocity vectors of the two clouds, respectively; $r_{\rm pe}$ and $r_{\rm
ap}$ are the pericentres and the apocentres of the two clouds; $e$ is the eccentricity
of their orbits; $\theta$ is the angle between the orbital planes of the
clouds, and $b$ is the impact parameter.}

\vspace{0.1in}
\begin{center}
\begin{tabular}{|c|c|c|c|c|c|c|c|c|c|c|c|}\hline
ID & $\beta$ & $\bf{v_{1}}$ & $\bf{v_{2}}$ &
                                  $r_{1,\rm pe}$ & $r_{1,\rm ap}$&
                                  $r_{2,\rm pe}$ & $r_{2,\rm ap}$ & $e_{1}$ & $e_{2}$ & $\theta$ ($^{\circ}$) & $b$\\ \hline
                                  \hline
S1 & 1  & (0,0.2,0) & (0,-0.11,-0.21) & 25  & 31.5 & 12.8 & 29.5 & 0.12 & 0.39 & 116 & 3.8\\ \hline
S2 & 1  & (0,0.2,0) & (0,-0.21,-0.11) & 25  & 31.5 & 13.2 & 29.1 & 0.12 & 0.38 & 151 & 6.8\\ \hline
S3 & 0.3 & (0,0.2,0) & (0,-0.11,-0.21) & 25 & 31.5 & 12.8 & 29.5 & 0.12 & 0.39 & 116 & 3.8\\ \hline
S4 & 0.3 & (0,0.2,0) & (0,-0.21,-0.11) & 25 & 31.5 & 13.2 & 29.1 & 0.12 & 0.38 & 151 & 6.8\\ \hline
S5 & 1 & (0,0.2,0) & (0.16,-0.11,-0.21) &25 & 31.5 & 21.8 & 43.1 & 0.12 & 0.33 & 120 & 2.4\\ \hline
S6 & 1 & (0,0.2,0) & (0.16,-0.21,-0.11) &25 & 31.5 & 21.5 & 43.4 & 0.12 & 0.34 & 147 & 5.3\\ \hline
\end{tabular}
\end{center}
\label{table1}
\end{table}

We run tests with cooling parameter $\beta = 1$ and $\beta = 0.3$ (see
equation \ref{beta}). These values are low enough so that fragmentation would
occur if and when regions of the gas became self-gravitating
\citep{Gammie01,Rice05}. Since the faster cooling runs were expected to
require on average shorter timesteps the accretion radius for $\beta = 0.3$
was set to $r_{\rm acc} = 0.33$ whilst for $\beta = 1$ this was set to $r_{\rm acc} = 0.06$.

\section{Numerical simulations}

We shall now present some of the results of our simulations. 
In all of our runs, the clouds undergo an off-centre collision at time
$t \sim 10$ ($\approx 600$ years). As the cooling time is longer than the
collision time, $t_{\rm coll} \sim (R_1 + R_2)/(|\bf{v_1} - \bf{v_2}|)$, the
clouds heat up significantly and hence initially expand due to thermal pressure. This thermal
expansion modifies velocities of the different parts of the clouds by
giving gas thermal velocity ``kicks''. The net result is a distribution of gas
velocities that is much broader than we would get if the shock were isothermal.\\

\noindent \textbf{Gas dynamics}

\noindent The collision's product is not self-gravitating (until it has cooled
down again) and so is easily sheared by the tidal field of \sgra. The
collision and the resulting mixing of the clouds leads to angular momentum
cancellation in shocks of some parts of the gas. Regions of gas that acquire
smaller angular momentum infall to the respective circularisation radius on
the local dynamical time. A small-scale disc around the black hole is thus
formed on this timescale. Regions of the clouds that did not directly
participate in the collision retain more of their
initial angular momentum. These regions are sheared, then cool into
filaments of length comparable to the initial sizes of the clouds'
orbits. Parts of these filaments collide with each other or with the inner
disc if the pericentres of their orbits are small enough. The inner disc
therefore gains mass in an asymmetric manner, possibly resulting in strong
warping and a change in its orientation with time.

As the mass of the disc increases, it circularises, cools and undergoes
gravitational collapse, with high density gas clumps being formed. Sink
particles are introduced inside these clumps and allowed to grow in mass via
gas accretion. We therefore see a significant amount of star formation, and in
all our simulations, although the precise distributions differ, we always end
up with at least two distinct stellar populations; in the disc and in the
outer stream(s).

\begin{figure}[ht]
\begin{minipage}[b]{.5\textwidth}
\centerline{\psfig{file=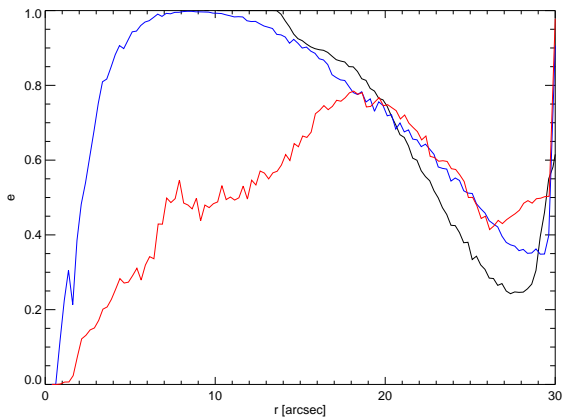,width=0.99\textwidth,angle=0}}
\end{minipage}
\begin{minipage}[b]{.45\textwidth}
\caption{ Gas eccentricity defined on radial shells as a function of the shell's
  radius at different times.  Black, blue and red curves corresponding to time
  $t=50$, 100 and 250. Note that initially only gas on near-plunging high
  eccentricity orbits arrives in the innermost region, but with time gas
  circularises to $e\ll 1$.}
\end{minipage}
\label{fig:ecc_radialplot}
\end{figure}

\noindent Figure 1 shows the profiles of orbital eccentricity defined
on radial shells plotted for several snapshots.  These profiles show that soon
after the collision the inner part of the computational domain is dominated by
gas on plunging -- high eccentricity -- orbits. At later times, shocks force
the gas to circularise at $r \simlt$ a few. Gas circularises faster at smaller
radii as the orbital time is shortest there. As time goes on, nearly circular gaseous orbits are established in the innermost
few arcseconds, whereas more eccentric ones dominate at larger radii.\\

\noindent \textbf{Star formation}

\noindent Stars in all of our runs form in both the disc and at least one
filament. In the disc, stars form first in the inner arcsecond at $t \sim 900$ and later at a radius of $\sim 5$-$8''$ at $t \sim 1700$. In the filament(s), star formation is approximately co-eval with the inner arsecond population and occurs at a radius of $\sim 15$-$25''$. The disc and filament stellar orbits are inclined to each other by $\sim 60^{\circ}$. Mass functions in the disc are top-heavy, whilst the filament(s) form entirely low-mass stars ($0.1$-$1 \msun$). However, as outlined
in \S \ref{sec:numerics}, the mass spectrum of stars formed depends on several
poorly constrained parameters of the simulations \citep[e.g.,
see][]{NayakshinEtal07}.  Indeed, in the faster cooling runs, the stars in the outer
filament(s) form mostly in a clustered mode. We expect that addition of
radiative transfer and feedback would supress artificial fragmentation in
these clusters and lead to a much more top-heavy mass function of the stars
formed in the filament(s).

\begin{figure*}
\begin{minipage}[b]{.50\textwidth}
\centerline{\psfig{file=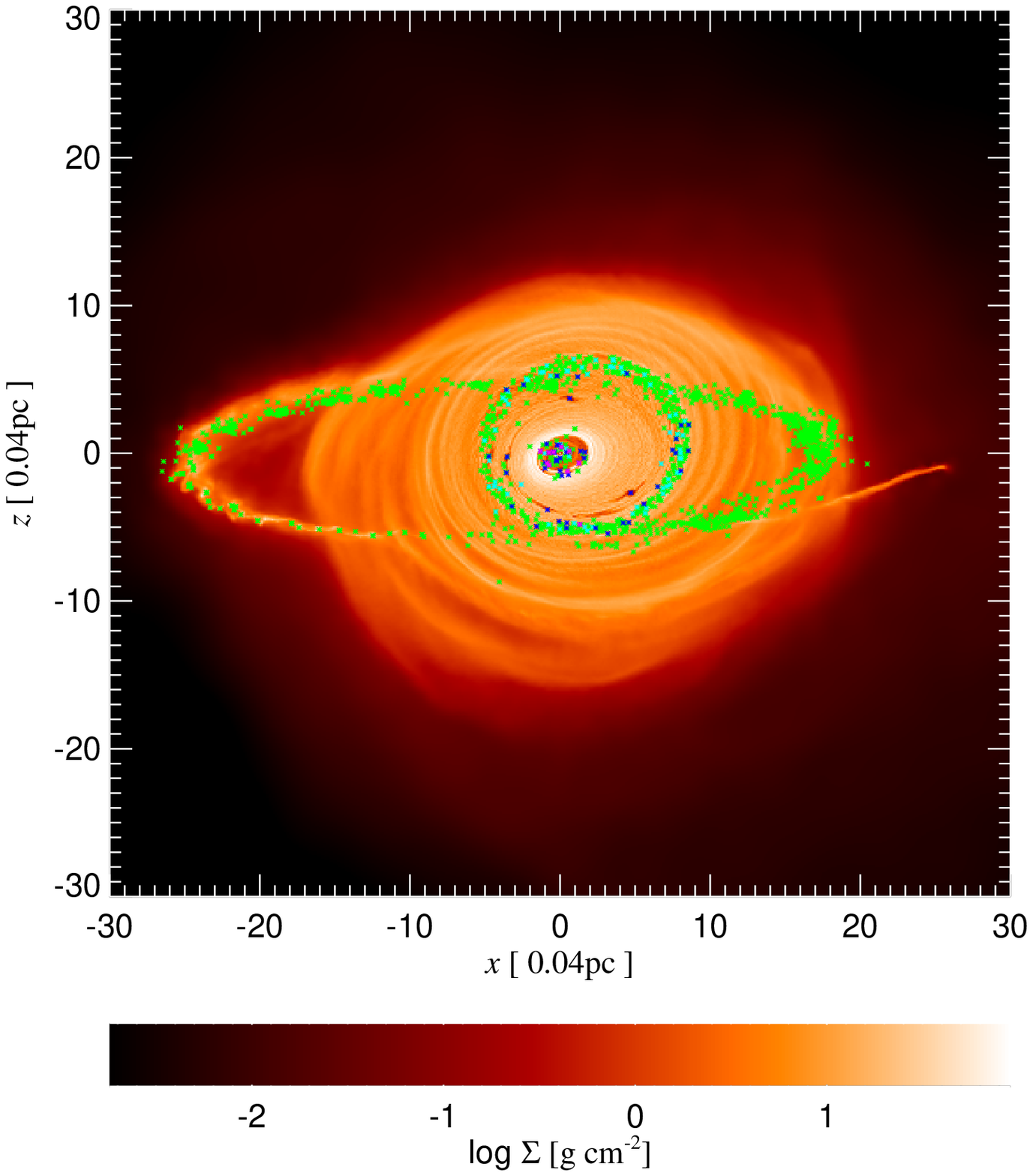,width=0.99\textwidth,angle=0}}
\end{minipage}
\begin{minipage}[b]{.50\textwidth}
\centerline{\psfig{file=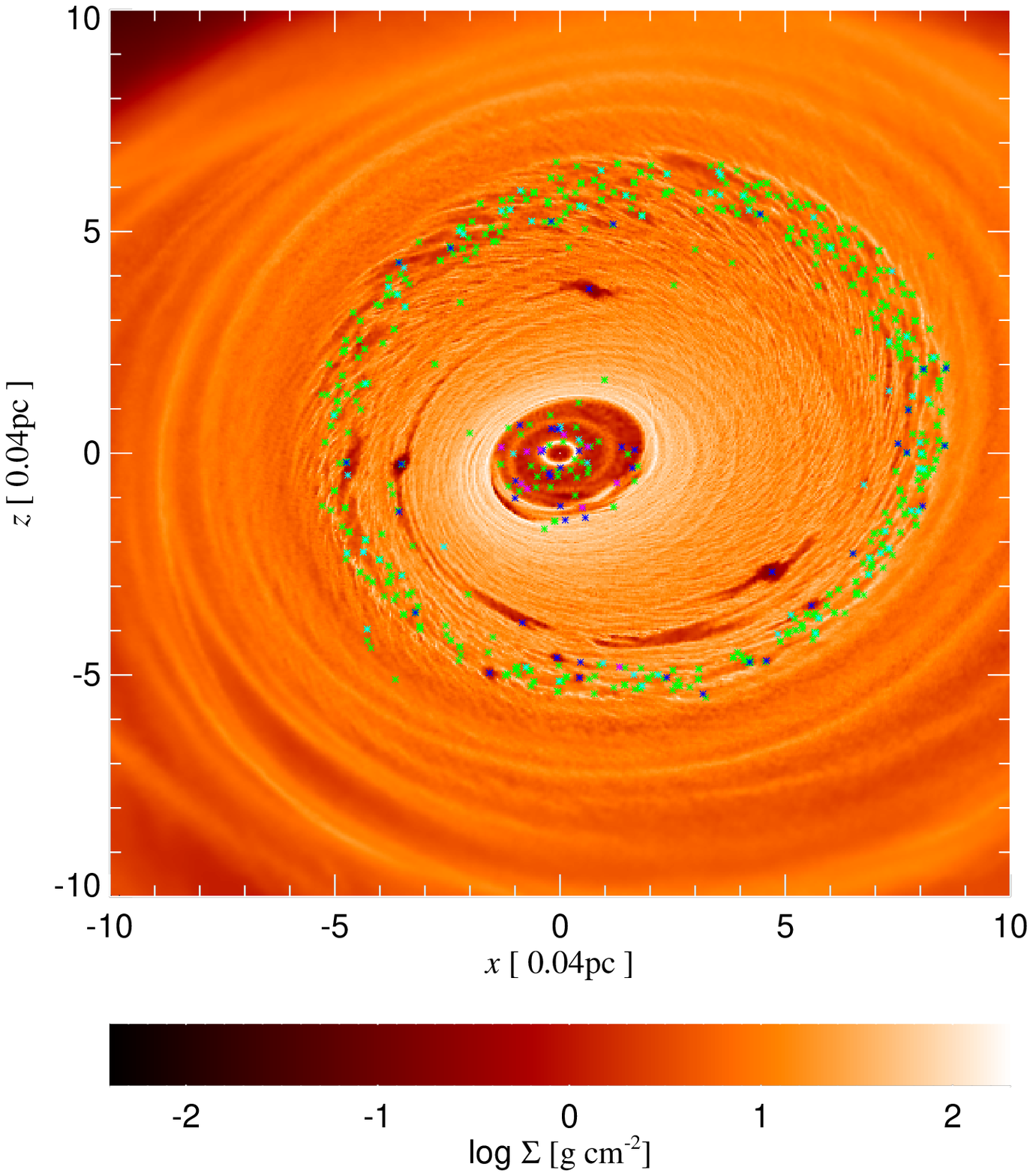,width=0.99\textwidth,angle=0}}
\end{minipage}
\caption{Gas surface density and stars for simulation S1 at time $t=1955$. Left: the whole of the central parsec, showing the inner disc, mid-range disc, and filament stellar populations. Right: a zoomed-in plot of the inner disc. \sgra\ is located at (0,0),
  and the line of sight is along the $y$ direction. Stars are colour-coded by
  mass: green (0.1-1 $\msun$), cyan (1-10 $\msun$), dark blue (10-150 $\msun$),
  magenta ($>$ 150 $\msun$).}
\label{fig:gasandstars}
\end{figure*}

All three stellar populations can be seen in Figure \ref{fig:gasandstars}. We distinguish the locations at which they form by ``inner disc'' (within the inner 1-2$''$), ``mid-range disc'' (5-8$''$) and ``filament'' (self-explanatory). The orbits in the inner disc population are almost circular, $e \sim 0.05$, whilst the mid-range disc population has an eccentricity of $e \sim 0.2$. Both of these are therefore in good agreement with the clockwise feature of the observations \cite{LuEtal06}. The orbit of the filament population is also eccentric, with $e \simgt 0.2$. This is in reasonable agreement with the corresponding counter-clockwise feature of the observations, although the eccentricity of the latter is considered to be significantly higher with $e \sim 0.8$ \cite{PaumardEtal06}.\\

\noindent \textbf{Infall of gas onto inner disc}

\noindent When the feeding into the inner 1-2$''$ is reasonably steady-state, and cooling
is long, stars form here once the gaseous inner disc has settled into a stable orientation.
The resulting stellar distribution of is that of a
geometrically thin disc ($H/R < 0.1$), a distribution
that is broadly consistent with the observed orbits of young massive stars in the
clockwise disc \citep{PaumardEtal06}.

In some of our runs, however, the inner disc undergoes a complex
evolution in angular momentum space. When the feeding of the disc due to
infalling gas is intermittent in its distribution of angular momentum e.g.
due to a more grazing initial collision, the newly arriving gas tweaks the disc orientation
significantly. However, if stars are introduced in the disc whilst its
midplane orientation is still changing (as is the case for the $\beta = 0.3$
runs), these stars remember the ``old'' orientation of the gaseous disc in
which they were born whilst the gaseous disc evolves to quite a
different orientation. Possibility of this effect taking place was suggested
based on analytical arguments by \cite{NC05}. These authors found a ``critical
rotation timescale'', below which the disc will leave the stars behind as its
midplane rotates. Indeed, in one of our tests with fast cooling and a grazing
initial collision, the inner disc rotates on a timescale much shorter than this
critical value, and this coupled with several star forming events in gaseous discs of different
orientations leads to a geometrically thick stellar disc ($H/R\sim 1$) in the inner arcsecond.\\

\noindent \textbf{Radial distribution of stars}

\begin{figure}
\begin{minipage}[b]{.6\textwidth}
\centerline{\psfig{file=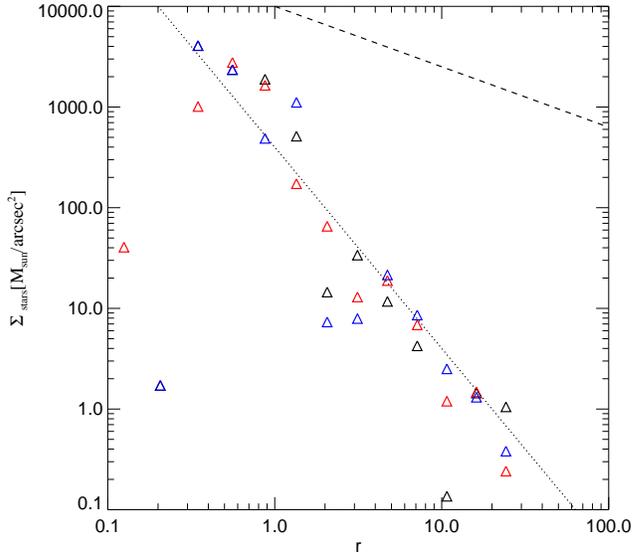,width=0.99\textwidth,angle=0}}
\end{minipage}
\begin{minipage}[b]{.35\textwidth}
\caption{Stellar surface density, $\Sigma_*$,  versus projected radius, for
  the run S1 at t$=1955$. Different colours show the three
  different orientations, along the $z$, $x$ and $y$ axes (black, red, and
  blue respectively).}
\end{minipage}
\label{fig:sigma_vs_r}
\end{figure}
\noindent Interestingly, the most robust result from all of the 6 runs completed is the
fact that distribution of stellar mass in an annulus, $dM_* = \Sigma_*(R) 2\pi
R dR$, versus radius roughly follows the law
\begin{equation}
\Sigma_*(R) \propto \frac{1}{R^2}\;.
\label{sigma_vs_r}
\end{equation}
We show these distributions for run S1 in Figure \ref{fig:sigma_vs_r}.  To
facilitate comparison with observational data, we use {\em projected} radius
$p$ instead of the proper 3D radius $r$, which is not directly known in
observations of young stars in the GC \citep[e.g.,][]{GenzelEtal03}. As the
viewing angle of the stellar system modelled here is arbitrary, we chose to
plot the distributions along the three axes of the simulations, so that $p =
\sqrt{x^2 + y^2}$ (black symbols), $p = \sqrt{y^2 + z^2}$ (red), and $p = \sqrt{x^2 + z^2}$ (blue). The lines in the
Figure show $R^{-2}$ (dotted) and $R^{-3/5}$ (dashed) power laws. The latter
corresponds to that predicted by a {\em not self-gravitating} standard
accretion disc theory.  This result is remarkable given very large differences
in the 3D arrangments of the stellar structures in our runs. The $1/R^2$ law
is in a good agreement with the observed distribution \citep{PaumardEtal06}.\\

\noindent \textbf{Sgr A* feeding during and after the cloud collision}

\begin{figure}
\begin{minipage}[b]{.60\textwidth}
\centerline{\psfig{file=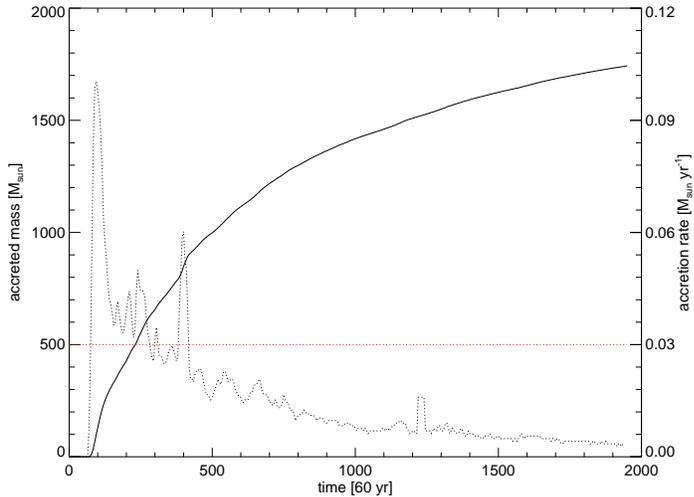,width=0.99\textwidth,angle=0}}
\end{minipage}
\begin{minipage}[b]{.40\textwidth}
\caption{The time evolution of the accreted mass onto the central black hole (solid line)
and the corresponding accretion rate (dotted line) compared to the Eddington limit for
Sgr A* (dotted red line), for simulation S1.}
\end{minipage}
\label{fig:accretionhistory}
\end{figure}

\noindent Figure \ref{fig:accretionhistory} shows the mass accretion rate for
the central black hole for simulation S1. Formally, our simulations provide a
sustained super-Eddington accretion rate ($\simgt 0.03 \msun$ yr$^{-1}$), in
some cases for over $10^{4}$ yrs, and at a fraction of that for up to $10^5$
years. However, we do not resolve gas dynamics inside the accretion radius,
where the material should form a disc in which accretion proceeds
viscously. The viscous time scale, $t_{\rm visc}$, depends on the temperature
in the disc midplane and the viscosity parameter, $\alpha$
\citep{Shakura73}. The midplane temperature in the inner arcsecond is around
$10^3 K$ in both the standard \citep[non self-gravitating, e.g.,][]{NC05} and
the self-gravitating regimes \citep{Nayakshin06a}, yielding $H/R \sim 0.01$:
$t_{\rm visc} = 6\times 10^6 \;\hbox{years}\; \alpha_{0.1}^{-1} (R/100H)^2  r^{3/2}\;,$
where $\alpha_{0.1} = \alpha/10$ and $r$ is in our code units (one arcsecond).
With these fiducial numbers, the viscous time scale coincides with the age of
the young massive stars in the GC \citep[e.g.,][]{KrabbeEtal95,PaumardEtal06}.
If $t_{\rm visc} \ll$ a few million years (e.g., if $\alpha = 0.1$, $t_{\rm visc}
\approx 10^6$ years at 0.3''), we expect that gas would have mainly accreted
onto \sgra\ by now. This accretion rate would be a significant fraction of the
Eddington accretion rate. Standard disc accretion would then generate as much
as $2\times 10^{56}$ erg $M_{\rm acc, 3}$ of radiative energy, where $M_{\rm
acc, 3}$ is the gas mass accreted by \sgra\ in units of $10^3 \msun$. A
similar amount of energy could have been released as energetic outflows. There
is currently no evidence for such a bright and relatively recent accretion
activity of \sgra.

In the opposite limit, i.e. if $\alpha \ll 0.1$, and $t_{\rm visc} \simgt$ a
few million years, the gaseous disc should still be there as self-consistent
modelling predicts that stars should not be forming within region $r \simlt
0.3''-1''$ \citep{Nayakshin06a}. This would contradict the data, as no gaseous
disc is observed now \citep{Falcke97,Cuadra04}.  Therefore, our
simulations seem to over-predict the amount of gaseous material deposited in
the inner $\sim 0.3''$. We take this as an indication that a better model
would perhaps involve a gaseous cloud of a larger geometric size, thus
shifting all spatial scales outwards.

\section{Discussion}\label{sec:discussion}

In approximate order of robustness, our key results are as follows: 

\begin{enumerate}

\item Formation of a gaseous nearly circular disc in the inner region of the
computational domain is common to all runs, as is the ensuing formation of stars
on similar circular orbits. This is natural as the dynamical time in the
innermost disc is only $\sim 60 r^{3/2}$ years. Conversely, the outer gaseous
stream becomes self-gravitating much faster than it could circularise, and
hence orbits of stars in that region are more eccentric, in reasonable agreement with 
observations \citep{PaumardEtal06}.

\item Radial distribution of stellar mass closely follows the
observed $\Sigma_* \propto 1/R^2$ profile of the disc stellar populations
\citep{PaumardEtal06}. This was observed for all the runs, although we expect
these results to change if colliding clouds moved in similar directions,
significantly reducing the angular momentum cancellation in shocks and the
thermal ``kick'' velocity due to the shock.

\item Runs with a comparatively long cooling time parameter, $\beta=1$, lead
to kinematically less dispersed stellar populations than runs with faster
cooling. As a result, the longer the cooling time, the more closely the
resulting stellar system can be fit by planar systems in velocity space. The
innermost stellar disc is then reminiscent of the observed clockwise {\em
thin} stellar system \cite{PaumardEtal06}.  Rapidly cooling runs produce
clumpier gas flows that lead to significant gaseous disc orientation changes,
producing less coherent discs; such geometrically thick systems are
incompatible with the observations.

\item With the chosen initial conditions, faster cooling promotes
survival of the gaseous streams corresponding to the orbits of the original
clouds. These streams fragment and form stars mainly in a clustered mode,
although this is expected to depend on the details of the radiation
feedback from young stars, which is not modelled in this set of simulations.

\end{enumerate}

\noindent The observed well defined, flat, geometrically thin and almost circular
clockwise stellar system \citep{PaumardEtal06} is best created via a gentle
accumulation of gas. Several independent major gas deposition events lead to a
too warped disc, and/or mixed up systems consisting of several stellar rings
or discs co-existing at the same radius. To avoid this happening, the inner
disc must be created on a timescale longer than the critical rotation time,
which is estimated at $t_{\rm cr} \sim$~few~$\times 10^4$~years.

Deposition of gas onto the inner disc will most likely take place over the
collision time, $t_{\rm coll} \sim R_{\rm cl}/v_{\rm cl}$, where $R_{\rm cl}$ and
$v_{\rm cl}$ are the cloud's size and velocity magnitude. We therefore require
that the collision itself be more
prolonged than $t_{\rm cr}$. Estimating the velocity of the cloud at $v_{\rm cl}
\sim 150$ km/sec, which is of the order of circular velocities in the GC
outside the inner parsec, we find $t_{\rm coll} = R_{\rm cl}/v_{\rm cl} \sim
10^4 \hbox{years}\; R_{\rm cl, pc}$, where $R_{\rm cl, pc}$ is the size of the
cloud in parsecs. We hence require the cloud to be larger than a few parsecs to
satisfy $t_{\rm coll}\simgt t_{\rm cr}$. Note that this size is not
necessarily the original size of the cloud if the cloud gets tidally disrupted
before it makes the impact. In the latter case we can take $R_{\rm cl}$ to be
the radial distance to the centre of the Galaxy at which the tidal disruption
took place. Finally, the location of the collision should not be too far from
the central parsec, or else too much angular momentum would have to be lost to
deposit a significant amount of gas in the $\sim 0.1$ pc region.   

In addition, the radial distribution of gas and stars in our simulations is
too compact, contradicting the observations (no He-I stars inside the inner
arcsecond and no gaseous disc there either). We thus believe that a realistic
scenario would be a GMC of the order of a few parsecs in size striking the CND 
a few parsecs away from \sgra.
Alternatively such a cloud could self-collide if the impact parameter with
respect to \sgra\ is small enough, but the cloud would need to be very structured,
e.g., essentially consist of several smaller clouds or filaments.

\bibliographystyle{iopart-num} 

\bibliography{nayakshinalex_jpconf}

\end{document}